\documentclass{emulateapj}
\begin{document}
\title{ The Cosmic M\lowercase{e}V 
Gamma-ray Background and Hard X-ray Spectra of
Active Galactic Nuclei: Implications for the Origin of Hot AGN Coronae}
\author{Yoshiyuki Inoue, Tomonori
Totani, and Yoshihiro Ueda}

\affil{Department of Astronomy, Kyoto University
, Kitashirakawa, Sakyo-ku, Kyoto 606-8502, Japan}
\email{yinoue@kusastro.kyoto-u.ac.jp}

\begin{abstract}

The origin of the extragalactic gamma-ray background radiation at
1--10 MeV is still unknown. Although the cosmic X-ray background (CXB)
up to a few hundreds keV can be accounted for by the sum of Active
Galactic Nuclei (AGNs), current models of AGN spectra cannot explain
the background spectrum beyond $\sim$1 MeV, because of the thermal
exponential cutoff of electron energy distribution assumed in the
models. Here we construct a new spectral model by calculating the
Comptonization process including nonthermal electrons, which are
expected to exist in an AGN hot corona if it is heated by magnetic
reconnections. We show that the MeV background spectrum can nicely be
explained by our model, when coronal electrons have a nonthermal
power-law component whose total energy is a few percent of the thermal
component and whose spectral index is $ d\ln N_e/d\ln E_e \approx
-4$. Although the MeV gamma-ray flux from such a component in nearby
AGN spectra is below the detection limit of past observations, it
could be detected by planned future MeV detectors. We point out that
the amount of the nonthermal component and its electron index are
similar to those found for electrons accelerated by magnetic
reconnections in solar flares and the Earth magnetosphere, giving a
support to the reconnection hypothesis for the origin of hot AGN
coronae.

\end{abstract}


\section{Introduction}

It is well-known that normal active galactic nuclei (AGNs) explain the
cosmic X-ray background (CXB) below several hundreds keV (for reviews
see {Boldt} 1987; {Fabian} \& {Barcons} 1992; {Ueda} {et~al.} 2003, hereafter U03; {Gilli}, {Comastri}, \&  {Hasinger} 2007). It is also known that rare AGNs of the blazar
type make a considerable contribution to the cosmic gamma-ray background
in the energy range from 100 MeV to 100 GeV, which has a hard power-law
spectrum (almost flat in the $\nu F_\nu$ plot), though blazars may not
explain all of the background flux, leaving some room for possible
contributions from completely different sources ({Narumoto} \& {Totani} 2006, and references
therein).

The origin of the gamma-ray background at the gap between these two
energy regions, i.e., $\sim$1--10 MeV, has also been an intriguing
mystery. The AGN spectra adopted in population synthesis models of the
CXB cannot explain this component because of the assumed exponential
cut-off at a few hundred keV. The background spectrum in the 1--10 MeV
band is much softer (photon index $\alpha \sim $2.8) than the GeV
component, indicating a different origin from that above 100 MeV
(e.g. {Sreekumar} {et~al.} 1998).

A few candidates have been proposed to explain the 1--10 MeV
background. One is the nuclear decay gamma-rays from type Ia supernovae
(SNe Ia) ({Clayton} \& {Ward} 1975; {Zdziarski} 1996; {Watanabe} {et~al.} 1999). However, recent studies based on the
latest measurements of the cosmic SN Ia rates show that the background
flux expected from SNe Ia is about an order of magnitude lower than
observed ({Ahn}, {Komatsu}, \&  {H{\"o}flich} 2005; {Strigari} {et~al.} 2005). There is
a class of blazars called ``MeV blazars'', whose spectra have peaks at
$\sim$ MeV ({Blom} {et~al.} 1995; {Sambruna} {et~al.} 2006), and these MeV blazars could potentially contribute to the MeV background. However, quantitative estimate of the contribution is difficult because
of the still small sample available at present. Annihilation of the dark matter particles
has also been discussed
({Ahn} \& {Komatsu} 2005a, 2005b; {Rasera} {et~al.} 2006; {Lawson} \& {Zhitnitsky} 2007), but there is no natural particle physics candidate for such a dark matter
particle with a mass scale of $\sim$ MeV.  The motivation of MeV dark
matter has been inspired by the 511 keV line emission from the Galactic
center, but a few astrophysical explanations are possible for this line
emission ({Totani} 2006, and references therein).

An important feature of the MeV background spectrum is that its
power-law spectrum is {\it smoothly} connected to the peak of the CXB
spectrum. If the origin of the MeV background is completely different
from that of the CXB, such a smooth connection would be
surprising. Rather, it seems more plausible that the MeV background
flux is composed of the same populations that make the CXB, and simply
the current AGN spectral models are not sufficient to describe the MeV
spectra. The X-ray AGN spectra are well described by the
Comptonization of seed UV photons by the hot coronal
electrons ({Zdziarski} {et~al.} 1994, 1995), and the cut-off at $\sim$100 keV
reflects the thermal energy distribution of the hot
electrons. Although the AGN spectra indeed show evidence for such a
cut-off ({Zdziarski} {et~al.} 1995; {Zdziarski}, {Poutanen}, \&  {Johnson} 2000), a small amount of additional non-thermal
electrons with a soft spectrum is sufficient to explain the MeV
background. Due to the limited sensitivity of current MeV gamma-ray
observations, the presence of such non-thermal components is not
strongly constrained even in the spectra of nearby brightest AGNs.
Furthermore, it is believed that coronae around accretion disks share
some common features with the solar corona (e.g., {Galeev}, {Rosner}, \&  {Vaiana} 1979), and
magnetic reconnection in AGN coronae is a good candidate for the
origin of hot electrons ({Liu}, {Mineshige}, \&  {Shibata} 2002). It is well known that particles
are accelerated to nonthermal energies by reconnections in solar
flares ({Shibata} {et~al.} 1995).

Here we construct a new model of the X/gamma-ray spectra of AGNs, by
calculating the Comptonization process by hot electrons having both
thermal and nonthermal components. We also calculate the CXB spectrum
based on our model with the latest knowledge of the cosmological evolution of
the AGN luminosity function, and determine the amount and spectrum of
the nonthermal electrons in AGN coronae to explain the MeV
background. We discuss the implied nature of nonthermal electrons in the
context of the reconnection heating scenario of the AGN coronae,
comparing our results with those found in the reconnections occurring in
the solar flares and the Earth magnetosphere.

{Rogers} \& {Field} (1991) and {Field} \& {Rogers} (1993)
presented an AGN spectral model that can explain the MeV background
spectrum by nonthermal relativistic electrons. However, their model only
considers the nonthermal component without a thermal component, and it
requires a \textit{lower} cut-off of $\gamma_e \sim 30$ in the nonthermal
component, which is difficult to interpret as they mentioned in their
paper. Our model considers both the thermal and nonthermal coronal
electrons whose spectra are smoothly connected to each other, which is a
natural extension of the popular AGN spectral models in the recent
literature.  {Stecker}, {Salamon}, \&  {Done} (1999) also discussed a
possibility that the MeV background is explained by nonthermal tails in
AGN spectra, quoting the spectrum of the Galactic stellar-mass black
hole candidate Cyg X-1. However, a physical model to explain the
nonthermal tail in an AGN spectrum was not presented.

Throughout this paper, we adopt
the cosmological parameters of
$(h_0,\Omega_m,\Omega_\lambda$)=(0.7,0.3,0.7).

\section{Model Description}

\subsection{the AGN Spectra with Nonthermal Coronal Electrons}

The main shape of X-ray AGN spectra is determined by Comptonization of
UV photons emitted from optically-thick accretion disks by hot electrons
in coronae. As in many previous studies, we consider a simple spherical
and uniform distribution of the coronal electrons. The seed UV photons
are injected at the center and then become X-ray photons when they
escape the coronal region after Comptonization. In addition to the hot
thermal electrons assumed in the conventional X-ray spectral models of
AGNs, we introduce higher energy nonthermal electrons in AGN coronae,
whose energy distribution is a power-law as $dN_e/dE_e \propto E_e
^{-\Gamma}$. We introduce the transition electron Lorentz factor
$\gamma_{\rm{tr}}$, corresponding to the transition electron energy $E_e
= m_e \gamma_{\rm tr}$ where the electron spectrum $dN_e/dE_e$ has the
same value for the thermal and nonthermal components. This $\gamma_{\rm
tr}$ is the lower limit of the Lorentz factor distribution of the
nonthermal component, and hence there are no nonthermal electrons at
$E_e < m_e \gamma_{\rm tr}$. We also set an upper bound as $\gamma_{u} =
10^5$, although this hardly affects our results if the maximum photon
energy well extends beyond 10 MeV.

We set the coronal temperature to be $kT_e=256$ keV and assume a
blackbody spectrum for UV seed photons from a cooler disk with $T_d =
$ 10 eV, following the conventional thermal models
(e.g. {Zdziarski} {et~al.} 1994). The degree of Comptonization is determined by
the optical depth for Thomson scattering, $\tau_T$.  We found that the
spectral photon index in the X-ray band, $\alpha_X$, becomes close to
that typically found in observed spectra ($\alpha_X \approx
1.9$, e.g.,  {Nandra} \& {Pounds} 1994; {Turner} {et~al.} 1997; {George} {et~al.} 1998) when we set $\tau_T =
0.24$. Therefore we use this value throughout this letter; this value
is also similar to those used in the conventional models. It should be
noted that $\alpha_X$ is hardly changed even if we introduce the
nonthermal electron component with an amount that is necessary to
explain the cosmic MeV background.
 
We then trace the Comptonization process using a Monte Carlo
method. The calculation method used here is mainly based on that in
{Pozdniakov}, {Sobol}, \&  {Siuniaev} (1977) and {Gorecki} \& {Wilczewski} (1984), but their original formalism in
the laboratory frame is not optimized for the ultra-relativistic
region. To calculate the scattering by high energy nonthermal
electrons more efficiently, we added a new formulation in the rest
frame of relativistic electrons based on {Corman} (1970).

The reflection of X-ray photons by cool, optically thick matter is
also an important feature of AGN X-ray spectra. We calculate this by
using the PEXRAV model ({Magdziarz} \& {Zdziarski} 1995) in the XSPEC package as done in
U03. Because of the limitation for the acceptable input spectrum in
PEXRAV, we use a power-law spectrum ($\alpha_X = 1.9$) plus an
exponential cutoff at $E = $ 500 keV, which is a good approximation of
the Comptonized spectrum only with the thermal electrons. The newly
added nonthermal electrons would change the spectrum significantly
only at $E \gtrsim$ 1 MeV, and hence this treatment is appropriate for
the reflection component which is important only at $\sim$1--100 keV.

\subsection{Calculation of the Cosmic Background Radiation}

We calculate the CXB spectrum by integrating our AGN spectral model in
the redshift and luminosity space, using the X-ray AGN luminosity
function of U03. Following the same formulation given in U03, we take
into account the absorption column density distribution ($N_{\rm H}$
function) and the contribution from Compton-thick AGNs. We confirm
that our main conclusion hardly change if instead we use a more recent
population synthesis model by {Gilli} {et~al.} (2007), as described below.


\section{Results}

\begin{figure}[t]
\centering 
\plotone{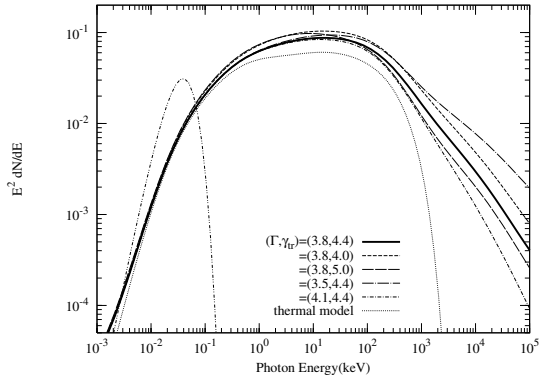} 

\caption{ The AGN spectra in X-ray and gamma-ray bands calculated by our
model. The flux is shown in an arbitrary unit of $E^2 dN/dE$, where
$dN/dE$ is a differential photon spectrum.  They are Comptonization of
UV seed photons without taking into account the reflection component and
the absorption effect. The thick solid curve is our standard spectrum
with $\Gamma=3.8$ and $\gamma_{\rm{tr}}=4.4$. The other thick curves are
for the cases of different model parameters as indicated in the
figure. The thick dotted curve is the spectrum only with the thermal
component ($kT_e=256$ keV).  The thin dotted curve is the input UV
spectrum (a black body with $T_d = 10$ eV).  \label{agn} }

\end{figure}

Figure \ref{agn} shows the models of AGN spectra calculated according
to the procedures in the previous section. Here, we do not take into
account the reflection component and the absorption effect, to show
the pure spectrum of the Comptonization. We set $\Gamma=3.8$ and
$\gamma_{\rm{tr}}=4.4$ as our standard model (solid line), because we
will find that these values give the best-fit MeV background spectrum
to the data.  In this standard model, $3.5\%$ of the total electron
energy is carried by the nonthermal electrons. To illustrate effects
of changing parameters, we also show the spectra with parameters
slightly changed from those for the best fit model.  The conventional
case only with the thermal electrons is also shown, where an
exponential cutoff above $\sim$100 keV is seen as expected.


\begin{figure}[t]
\centering 
\plotone{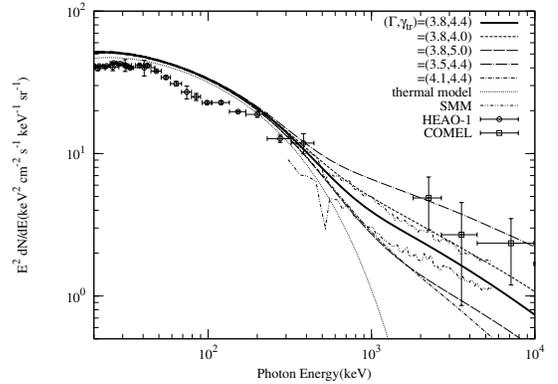} 
\caption{ The spectrum of the cosmic background radiation in X-ray and
gamma-ray bands, predicted by our model of AGN spectra shown in Fig.
\ref{agn}.  For each line-marking, the corresponding AGN spectrum in
Fig. \ref{agn} is used for the calculation.  The data points of HEAO-1
({Gruber} {et~al.} 1999) SMM
({Watanabe} {et~al.} 1999), and COMPTEL ({Kappadath} {et~al.} 1996) experiments are also shown. For the SMM data the thin dotted line indicates the $\pm 1 \sigma$ uncertainty region.
\label{egrb}}
\end{figure}

Figure~\ref{egrb} shows the cosmic background radiation integrated
with the luminosity function of U03 using our AGN spectral model. It
has been known that the predicted CXB spectrum below 100 keV by the
U03 model is 10--20\% higher than the HEAO-1 data. The origins of this
discrepancy are still controversial. The intensity and shape of the
CXB in the 10 keV -- 1 MeV band could be uncertain at $\approx$20\%
level\footnote{In Figure~2, a discontinuity is seen between the HEAO-1
A4 MED data above 100 keV and the HEAO-1 A4 LED
data below 100 keV ({Gruber} {et~al.} 1999, see also ; {Revnivtsev} {et~al.} 2005; {Churazov} {et~al.} 2007; {Frontera} {et~al.} 2007). If we use the model by {Gilli} {et~al.} (2007), the
discrepancy is reduced below 100 keV but is enhanced above 100
keV.}. The population synthesis models have also uncertainties, such
as the intensity of the reflection component assumed in the AGN
spectra, the (unknown) number density of Compton-thick AGNs, and the
parameters of the luminosity function itself that could be subject to
cosmic variance in deep surveys. However, here we emphasize that the
uncertainty hardly affects our conclusions. To confirm this, we also
calculate the MeV gamma-ray background flux with the U03 luminosity
function by artificially lowering its normalization by 20\% to match
the HEAO-1 data. We find that the best fit value of $\gamma_{\rm tr}$
is slightly changed by 0.4, while the change of $\Gamma$ is
negligible.

As mentioned above, we find here that the cosmic background spectrum
from X-ray to 10 MeV band can nicely be explained by our model with
$\Gamma = 3.8$ and $\gamma_{\rm tr} = 4.4$. The next question is then
how natural are these parameters in the context of the
theoretical picture of hot electrons in AGN coronae. We will discuss
about this issue in the next section.


\section{Discussions}
\subsection{Reconnection and Nonthermal Electrons}

As discussed in \S 1, magnetic reconnection is the primary candidate for
the origin of the nonthermal electrons in AGN coronae, and we can
compare the inferred amount and spectrum of electrons with those
observed in reconnections in other objects. In solar flares, the
electron spectrum accelerated by reconnection and injected into the
solar surface (foot points) can be estimated by thick-target
bremsstrahlung ({Brown} 1971; {Piana} 1994), and a value of $\Gamma \sim 4$
is inferred for giant solar flares ({Holman} {et~al.} 2003; {Lin} 2006).  In the
reconnection diffusion region of the Earth's magnetotail, $\Gamma = 3.8$
has been measured by the Wind spacecraft ({{\O}ieroset} {et~al.} 2002).  Interestingly,
these values are very similar to what we found to explain the MeV
background spectrum by AGNs.

The relative amounts of the thermal and nonthermal electrons are
difficult to predict, but it should be determined by the balance between
the cooling of the thermal electrons, thermalization of nonthermal
electrons, and energy input rate by reconnections. In solar flares, the
total energy of accumulated nonthermal electrons is comparable with or
even larger than that of the thermal electrons ({Holman} {et~al.} 2003).  In the
directly observed electron spectrum in the Earth's magnetotail, the
nonthermal component is smoothly connected to the thermal component at
the energy where the thermal electron spectrum declines by the
exponential cutoff, which is reminiscent of the cosmic background
spectrum from X-ray to the MeV band. Therefore, the inferred amount of
nonthermal electrons in AGN coronae seems quite reasonable in comparison
with the reconnections observed in the Sun or the Earth magnetosphere.

As in the solar flare, some of the nonthermal electrons in AGNs would be
injected into accretion discs along the magnetic field lines, and would
produce bremsstrahlung emission. Such emission may also contribute to
the AGN spectra. However, we consider here the inverse-Compton
scattering since X-ray AGN spectra are well explained by the
Comptonization process, indicating that the efficiency of
Comptonization is much higher than bremsstrahlung in AGNs, in contrast
to the case of X-ray radiation from solar flares. It should be noted
that the radiation efficiency of the thick-target bremsstrahlung is
generally much lower than the unity because the electrons injected into
dense material would lose their energy by ionization or Coulomb losses
rather than bremsstrahlung.

\subsection{Implications for Future MeV Observations of AGNs}

Our results suggest that the majority of AGN populations that are
responsible for the CXB should commonly have a nonthermal component
beyond the thermal exponential cutoff in their gamma-ray
spectra. Future sensitive MeV observations may reveal the nonthermal
components from nearby AGN spectra. For instance, the Advanced Compton
Telescope (ACT, see {Boggs} (2006)), scheduled for launch around
2015, will have a continuum sensitivity in the 0.2--10 MeV band with
$E_\gamma^2 dF_\gamma/dE_\gamma \sim 1 \times 10^{-5} (E_\gamma/{\rm
MeV}) \ \rm{MeV \ cm^{-2}s^{-1}}$, where $dF_\gamma/dE_\gamma$ is the
differential photon flux. Using the hard X-ray flux of NGC
4151 ({Sazonov} {et~al.} 2007), the brightest Seyfert galaxy, we estimate the
expected nonthermal MeV flux as $\sim 3 \times 10^{-5} (E_\gamma/{\rm
MeV})^{-0.8} \ \rm{MeV \ cm^{-2} s^{-1}}$, which can well be detected
with the ACT. The detection of such components from nearby bright AGNs
would give a clear test for our model.


\section{Conclusions}

In this paper, we have shown that MeV gamma-ray background can be
explained by the same population of AGNs that makes the CXB, by
considering Comptonization by nonthermal electrons in AGN coronae that
are theoretically expected to exist. The best fit to the MeV gamma-ray
background is obtained when the nonthermal component has $\sim 3.5\%$
of the total electron energy with a spectrum of $dN_e/dE_e \propto
E_e^{-3.8}$. This power law index is close to that of electrons
accelerated by magnetic reconnections in solar flares or the Earth
magnetosphere ({Holman} {et~al.} 2003; {Lin} 2006; {{\O}ieroset} {et~al.} 2002). This gives a support to
the idea of the reconnection-heated AGN corona ({Liu} {et~al.} 2002). There is
a chance for future MeV detectors with an improved sensitivity to
directly detect the nonthermal component predicted by our model from
nearby bright AGNs.

\acknowledgments
 
We would like to thank Eiichiro Komatsu for useful discussions, and
Duane Gruber and Ken Watanabe for providing their observational
data. This work was supported by the Grant-in-Aid for the 21st Century
COE "Center for Diversity and Universality in Physics" from the Ministry
of Education, Culture,Sports, Science and Technology (MEXT) of Japan.

\bibliography{}

\end{document}